\documentclass[twocolumn,aps,superscriptaddress,showpacs,nofootinbib,floatfix]{revtex4}
\usepackage{epsfig,bm,feynmf}
\usepackage{graphicx}
\usepackage{amsmath}
\usepackage{dcolumn}

\usepackage{graphicx}
\usepackage[usenames,dvipsnames,svgnames,table]{xcolor}

\usepackage{tikz-feynman,contour}
\tikzfeynmanset{compat=1.0.0}
\usepackage[normalem]{ulem}  

\renewcommand\sout{\bgroup \color{red} \ULdepth=-.5ex \ULset}

\begin{document}


\title{Bottomonium production in pp and heavy-ion collisions}


\author{Taesoo Song}\email{t.song@gsi.de}
\affiliation{GSI Helmholtzzentrum f\"{u}r Schwerionenforschung GmbH, Planckstrasse 1, 64291 Darmstadt, Germany}

\author{Joerg Aichelin}\email{aichelin@subatech.in2p3.fr}
\affiliation{SUBATECH UMR 6457 (IMT Atlantique,  Universit\'{e} de Nantes, IN2P3/CNRS), 4 Rue Alfred Kastler, F-44307 Nantes, France}
\affiliation{Frankfurt Institute for Advanced Studies, Ruth-Moufang-Strasse 1, 60438 Frankfurt am Main, Germany}

\author{Jiaxing Zhao}\email{jzhao@subatech.in2p3.fr}
\affiliation{SUBATECH UMR 6457 (IMT Atlantique, Universit\'{e} de Nantes,
	IN2P3/CNRS), 4 Rue Alfred Kastler, F-44307 Nantes, France}

\author{Pol Bernard Gossiaux}\email{Pol-Bernard.Gossiaux@subatech.in2p3.fr}
\affiliation{SUBATECH UMR 6457 (IMT Atlantique, Universit\'{e} de Nantes,
	IN2P3/CNRS), 4 Rue Alfred Kastler, F-44307 Nantes, France}

\author{Elena Bratkovskaya}\email{E.Bratkovskaya@gsi.de}
\affiliation{GSI Helmholtzzentrum f\"{u}r Schwerionenforschung GmbH, Planckstrasse 1, 64291 Darmstadt, Germany}
\affiliation{Institute for Theoretical Physics, Johann Wolfgang Goethe Universit\"{a}t, Frankfurt am Main, Germany}
\affiliation{Helmholtz Research Academy Hessen for FAIR (HFHF),GSI Helmholtz Center for Heavy Ion Research. Campus Frankfurt, 60438 Frankfurt, Germany}


\begin{abstract}

We study bottomonium  $b\bar b$ production  in pp collisions as well as in heavy-ion collisions, using a quantal density matrix approach. The initial bottom (anti)quarks are provided by the PYTHIA event generator. We solve the Schrödinger equation for  the $b\bar b$ pair, identifying the potential with the free energy, calculated with lattice QCD, to obtain the temperature dependent $b\bar b$ density matrix  as well as the dissociation temperature. The formation of bottomonium is given by projection of the bottomonium density matrix onto the density matrix of the system.  With this approach we describe the rapidity and transverse momentum distribution of the $\Upsilon $(nS) in pp collisions at $\sqrt{s_{\rm NN}}=$ 5.02 TeV extending a similar calculation for the charmonium states \cite{Song:2017phm}.  We employ the Remler formalism to  study the $b\bar b$ production in heavy ion collisions in which the heavy quarks scatter elastically with partons from the quark gluon plasma (QGP). The elastic scattering of heavy (anti)quark in QGP is realized by the dynamical quasi-particle model (DQPM) and the expanding QGP is modeled by PHSD.  We find that a reduction to 10 \% of the scattering cross section for a (anti)bottom quark with a QGP parton reproduces the experimental data. This suggests that due to color neutrality  the scattering cross section of the small  $b\bar b$ system with a parton is considerably smaller than twice the bottom-parton scattering cross section.

\end{abstract}


\maketitle
\section{introduction}

The conjecture of Matsui and Satz that the suppression of $J/\psi$ mesons, a colorless and flavourless bound state made up of a $c\bar c$ pair, is a signature for the formation of a QGP in heavy-ion collisions~\cite{Matsui:1986dk}, created wide research activities on quarkonium production in proton-proton and in heavy-ion collisions.

The production of a quarkonium involves perturbative and non-perturbative processes. The production of the heavy quarks requires a large momentum transfer and is therefore accessible to perturbative QCD.  The formation of the quarkonium out of a heavy quark and a heavy antiquark is non-perturbative. The typical momentum scale is the relative momentum between the quarks, which is of the order of 1 GeV.  Therefore
almost all models factorize the quarkonium production into a perturbative description of the production of the quark antiquark pair and a non-perturbative description of the formation of the bound pair. 
The perturbative part is usually formulated in terms  of leading order and/or next to leading order pQCD matrix elements. They include the leading order pair production and the next to leading order  gluon excitation and  gluon splitting processes and have been realized numerically by event generators like PYTHIA and EPOS.
Due to a lack of available guidance from the fundamental QCD, several models  have been advanced to describe the non-perturbative part. They include the Colour Octet Model (COM)
\cite{Braaten:1994kd,Braaten:1995cj,Bodwin:1994jh}, the Color Evaporation Model  (CEM) \cite{Amundson:1996qr} and the Color Singulet Model (CSM) \cite{Chang:1979nn}.  A short summary of these models can be found in Ref.~\cite{Andronic:2015wma}. The effective non-relativisitic QCD approach (NRQCD)  \cite{Caswell:1985ui,Bodwin:1994jh}  offers also the possibility to describe the transition from a quark-antiquark pair into a color singlet or color octet state and the transition between them,  governed by long distance matrix element, which have to be determined independently by fitting experimental cross sections.

Recently we have advanced a new idea to describe the non-perturbative part of quarkonium production \cite{Song:2017phm}.  It is based on the non-relativistic density matrix formalism. 
In this article we apply this formalism, which contains no free parameters, to the production of bottomonium and its excited states. 


Because the binding of a quarkonium is much stronger than that of light hadrons, hadronic interactions hardly  provoke the disintegration $J/\psi +h\to D +\bar D$ and therefore the $J/\psi$ multiplicity in heavy-ion collisions without the formation of a QGP would be approximately that produced in the initial hard collisions between the incoming nucleons (if one neglects the possible hadronic production, $D\bar D \to  J/\psi$). If, however,  a QGP is formed, then according to Matsui and Satz the color screening increases with increasing temperature  due to the colored partons in the QGP environment. This weakens the binding and therefore a disintegration $J/\psi \to c \bar c$ is possible, leading to a suppression of  $J/\psi$. Such a suppression has indeed been measured, first in the experiments at the Super Proton Synchrotron (SPS)~\cite{NA50:2004sgj,PHENIX:2006gsi} and later also in the Relativistic Heavy Ion Collider (RHIC) experiments.
However, if the collision energy increases, the cross section for heavy flavor production also increases.
If heavy quarks are produced abundantly in heavy-ion collisions, the possibility emerges that a heavy quark and a heavy antiquark from two different primordial  nucleon-nucleon collisions form a bound state when the temperature falls below the dissociation temperature of quarkonium. This contribution to the yield  is called the off-diagonal regeneration. It leads to an enhancement  of  the quarkonium multiplicity, which is opposite to the original idea of Matsui and Satz~\cite{Matsui:1986dk}.
Indeed, the nuclear modification factor of $J/\psi$ is larger at the Large Hadron Collider (LHC) than at the RHIC, though the collision energy is one order of magnitude higher at LHC than at RHIC and also the nuclear modification factor, $R_{\rm AA}$, is larger at midrapidity compared to  forward/backward rapidities, though temperature is higher at midrapidity~\cite{Manceau:2012ka}.
This observation is considered as  evidence for the regeneration of quarkonia in heavy-ion collisions.

There have been many different phenomenological models to describe quarkonium production in heavy-ion collisions~\cite{Andronic:2006ky,Grandchamp:2002wp,Yan:2006ve,Song:2011xi,Yao:2020xzw,Brambilla:2016wgg,Blaizot:2018oev,Villar:2022sbv,Ferreiro:2018wbd}.
The most critical quantities for an adequate description are the dissociation temperature and 
the energy loss in a QGP. If the dissociation temperature is low  most quarkonia will be dissolved in a QGP and only recombined quarkonia will be measured in experiment, as assumed in the statistical model~\cite{Andronic:2006ky}.
On the other hand, if the dissociation temperature of quarkonium is very high, both primordial and recombined quarkonia will be measured, as claimed in the two-component models~\cite{Grandchamp:2002wp,Yan:2006ve,Song:2011xi}. The potential between heavy quark and antiquark in a QGP is therefore crucial to understand the experimental data in heavy-ion collisions~\cite{Satz:2005hx,Petreczky:2018xuh,Lafferty:2019jpr}.  Knowing the potential, a Schrödinger equation can be employed to calculate the eigenstates and the binding energy of the quarkonium states.
The form of the potential as a function of the temperature, extracted from lattice calculations is, however,
still controversial~\cite{Lafferty:2019jpr, GaurangParkar:2022aoa}.

Being produced in a hard process the primordial heavy quarks have no preference in azimuthal direction.
This is also true for the quarkonia and therefore their initial  elliptic flow is zero. During the interaction
with QGP partons the heavy quarks may pick up the elliptic flow of the light partons, caused by the initial eccentricity of the interaction region in coordinate space. Therefore, the recombined quarkonia may have a finite elliptic flow. Consequently, the elliptic flow of quarkonia will allow to identify the relative importance of initial production and recombination~\cite{Liu:2009gx,Song:2010er}. 

The passage of heavy quarks through the QGP medium can be studied using the Remler formalism
\cite{Remler:1975fm,Remler:1975re,Remler:1981du}. The first calculations with this approach for the $J/\psi$  yield in an expanding QGP have been reported in Ref.~\cite{Villar:2022sbv}. This method does not separate primordial and recombined quarkonia but treats them simultaneously by calculating the expectation value of quarkonium density operator in the QGP in which heavy (anti)quarks  scatter with the light constituents of the QGP. In each of these collisions  dissociation and recombination may occur, depending on the momentum change of the heavy (anti)quark. The validity of this approach has been tested in a thermalized box with an initial out-of-equilibrium distribution of the heavy quarks in momentum and/or coordinate space \cite{Song:2023ywt}.

In this study we apply this method  for the study of bottomonium production in real heavy-ion collisions, where the expanding QGP is modelled by the Parton-Hadron-String dynamics (PHSD).  The comparison of the results of open heavy flavour mesons as well as that of light hadrons with experiments
\cite{Cassing:2008sv,Cassing:2009vt,Bratkovskaya:2011wp,Linnyk:2015rco}
has demonstrated that in PHSD the dynamics of heavy and light hadron is quite reasonably described in a wide range of collision energies~\cite{Song:2015sfa,Song:2015ykw,Song:2016rzw,Song:2018xca}.

This paper is organized as follows:
in Sec.~\ref{potential} we study the thermal properties of bottomonia in a QGP by solving the Schr\"{o}dinger equation with a temperature dependent heavy-quark potential from lattice QCD.
Section~\ref{pp} presents how bottomonium production is realized in pp collisions, revisiting our previous work on charmonium production in pp collisions~\cite{Song:2017phm}.  
Then the Remler's formula is applied to study bottonium production in heavy-ion collisions in Sec.~\ref{AA} and a summary is given in Section~\ref{summary}.

\section{heavy quark potential}\label{potential}

In heavy-ion collisions quarkonium production and disintegration is embedded in the hot environment of the QGP,  composed of partons. Therefore it is necessary to know the quarkonium properties in such a medium.  They depend on the potential between a pair of heavy quarks. We start out with the calculation of the temperature dependence of the eigenstates $\Phi_i$ of the heavy quark-antiquark pair. Because the
mass of the heavy quark is large as compared to their kinetic energy in the pair rest system we can use the non-relativistic Schrödinger equation. 

If the heavy quark potential is strongly attractive and therefore the dissociation temperature $T_{diss}$ is high, primordial quarkonia can survive the passage through the QGP.  Otherwise they will melt and most quarkonia measured in experiment are those which are regenerated late, when the temperature of the plasma falls below $T_{diss}$ .

No consensus has been reached about the most suitable heavy quark potential ~\cite{Satz:2005hx,Petreczky:2018xuh,Lafferty:2019jpr}. We assume, however, that the conclusions made in this paper do not depend on the specific choice of the potential. The  potential,  which we use,  is the free energy of a system made up of a heavy quark pair~\cite{Gubler:2020hft,Lee:2013dca}.
The free energy potential, extracted from lattice QCD calculations using the Wilson loop, assumes an infinite heavy quark mass. The relativistic corrections can be separated into two parts: the kinematic correction and  the dynamic correction caused by spin-spin interactions. The kinematic correction can be absorbed in the effective potential $V_{\rm eff}=V(r)-p^4/m_Q^3$ with $m_Q$ and $p$ being heavy quark mass and relative momentum, respectively.
So the relativistic correction leads to a deeper potential, the quarkonium becomes a more deeply bound state and the dissociation
temperature is enhanced. For the spin related term, we averaged over the different spin states. 
In the present study, however, we neglect both relativistic corrections. They are small due to the large mass of the bottom quark. The kinetic correction is $\propto 1/m_q^3$ , the spin-spin interaction is $\propto 1/m_q^2$. 

The free energy can be  decomposed into~\cite{Satz:2005hx} 
\begin{eqnarray}
F(r,T)=F_c(r,T)+F_s(r,T),
\label{eq:pot}
\end{eqnarray}
where $F_c$ and $F_s$, respectively, denote the Coulomb-like potential and the string potential. They are expressed as
\begin{eqnarray}
F_c(r,T)&=&-\frac{\alpha}{r}[e^{-\mu r}+\mu r],\nonumber\\
F_s(r,T)&=&\frac{\sigma}{\mu}\bigg[\frac{\Gamma(1/4)}{2^{3/2}\Gamma(3/4)}-\frac{\sqrt{\mu r}}{2^{3/4}\Gamma(3/4)}K_{1/4}[(\mu r)^2]\bigg],\nonumber\\
\end{eqnarray}
with $\alpha=\pi/12$, $\sqrt{\sigma}=0.445~{\rm GeV}$.  $\mu$ is the screening mass extracted from lattice calculations as a function of the temperature and parameterized as~\cite{Gubler:2020hft}:
\begin{eqnarray}
\frac{\mu}{\sqrt{\sigma}}&=&0.35+0.0034 \exp[(T/T_c)^2/0.22]~~~{\rm for~} T<T_c,\nonumber\\
\frac{\mu}{\sqrt{\sigma}}&=&0.45+0.5\bigg(\frac{T}{T_c}\bigg) \tanh\big[(T/T_c-0.93)/0.15\big] \nonumber\\
&&~~~{\rm for~} T>T_c,
\label{param}
\end{eqnarray}
At $T_c$ the two separate parametrizations match smoothly  with an (almost) identical derivative.



The properties of a quarkonium in the QGP can be obtained by solving the Schr\"{o}dinger
equation with the heavy quark potential:
\begin{eqnarray}
\bigg[2m_Q-\frac{1}{m_Q}\nabla^2+F(r,T)\bigg]\Phi(r,T)=M \Phi(r,T),\label{schrodinger1}
\end{eqnarray}
where $m_Q$ and $M$ are, respectively, heavy quark and quarkonium masses and $\psi(r,T)$
is the quarkonium wave function at temperature $T$. To reproduce the mass of $\Upsilon$ (1S) in vacuum
$m_Q$ is taken to be 4.62 GeV for the bottom quark.

\begin{figure}[h]
\centerline{
\includegraphics[width=9. cm]{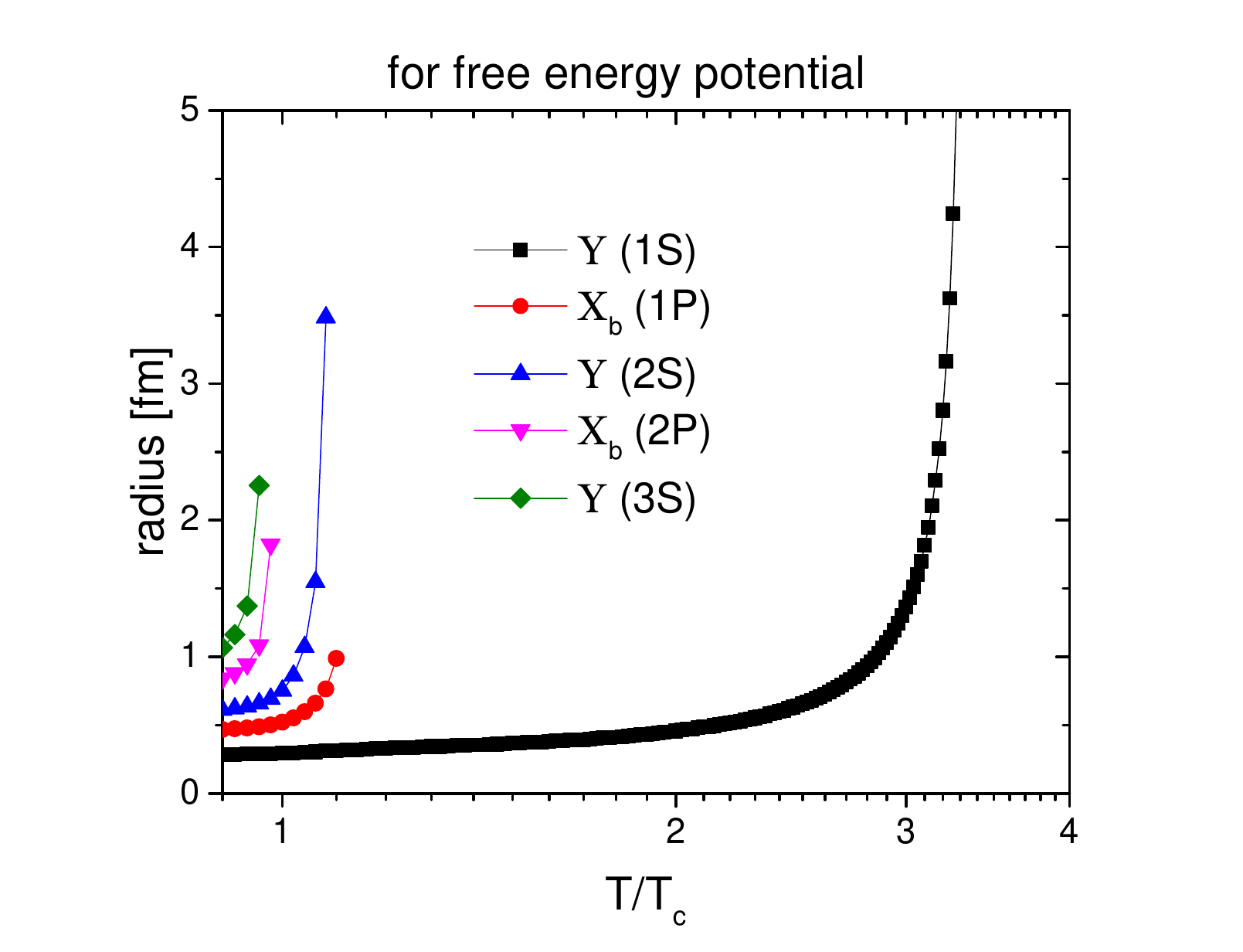}}
\caption{Radii of bottomonia as a function of temperature for the free energy potential.}
\label{radii}
\end{figure}

Figure~\ref{radii} shows the radii, $(r_Q-r_{\bar{Q}})/2$, of bottomonia as a function of the temperature.  They  are obtained by solving Eq.~(\ref{schrodinger1}) with the free energy potential of Eq.~(\ref{eq:pot}).
The dissociation temperature of $\Upsilon$ (1S) is about 3.3 $T_c$ while that of the excited states is less than 1.2 $T_c$.
Bottomonium radii slowly increase with temperature and then rapidly change near their dissociation temperatures.  This temperature-dependent radius of each bottomonium state will be used in the calculation presented later.
\section{Quarkonium Production in proton-proton collisions}\label{pp}

   A partial Fourier transformation converts the  density matrix of the eigen states  $|\Phi_i\rangle\langle\Phi_i|$  into a Wigner density 
\begin{equation}
\Phi^W_i({\bf r},{\bf p})=\int d^3y e^{i{\bf p\cdot y}}\langle{\bf r}-\frac{1}{2}{\bf y}|\Phi_i\rangle\langle\Phi_i|{\bf r}+\frac{1}{2}{\bf y}\rangle.
\label{eq:jpsiwign}
\end{equation}
normalized to $ \int \frac{d^3rd^3p}{(2\pi)^3}\ \Phi^W_i({\bf r},{\bf p})=1$. 
We assume for the moment that only one heavy quark pair is produced in these collisions to which we assign the coordinates  $\bf{r_1, p_1, r_2, p_2}$.  Employing the relative and center of mass coordinates
\begin{eqnarray}
{\bf R}=\frac{{\bf r_1}+{\bf r_2}}{2},~~~{\bf r}={\bf r_1}-{\bf r_2},\nonumber\\
{\bf P}={\bf p_1}+{\bf p_2},~~~{\bf p}=\frac{{\bf p_1}-{\bf p_2}}{2},~~
\label{define1}
\end{eqnarray}
we can calculate (neglecting all flavour, color and spin factors) the probability that a quarkonium eigenstate $i$ with momentum ${\bf P}$ is produced at ${\bf R}$:
\begin{eqnarray}
n_i({\bf R,P}) &=&\sum \int \frac{ d^3r d^3p}{(2\pi)^3}\  \Phi^W_i({\bf r},{\bf p}) \prod_{j> 2} \int \frac{ d^3r_j d^3p_j}{(2\pi)^{3(N-2)}} \nonumber \\
&& W^{(N)}({\bf r_1,p_1,r_2,p_2,...,r_N,p_N}),
\label{eq:ni}
\end{eqnarray}
where $W^{(N)}({\bf r_1,p_1,r_2,p_2,...,r_N,p_N})$ is quantal density matrix in Wigner representation of the $N$ partons produced in a proton-proton or a heavy-ion collision. Among them is a variable number of quark-antiquark pairs. 
The $\sum$ on the right hand side of Eq.~(\ref{eq:ni}) means the sum over all possible heavy quark-antiquark pairs among the $N$ partons, which are arranged in a way that the selected heavy quark - antiquark pair carries always the index 1 and 2. 
The probability that a quarkonium $i$  is formed is given by
\begin{equation}
P_i=\int \frac{ d^3Rd^3P}{(2\pi)^3} n_i({\bf R,P})
\end{equation}
whereas
\begin{equation}
  \frac{dP_i}{d^3P}=\int \frac{ d^3R}{(2\pi)^3} n_i({\bf R,P})
\label{eq:dnd3p}  
\end{equation}
is its momentum distribution. The quantal density matrix $W^{(N)}({\bf r_1,p_1,r_2,p_2,...,r_N,p_N})$ is not known. We assume here, and this is the basis of all transport approaches for heavy ion collisions, that we can replace the quantal N-body Wigner density by
the average of classical phase space distributions $W^{(N)}_{class}$ 
\begin{equation}
    W^{(N)}\approx \langle W^{(N)}_{class}\rangle.
    \label{eq:class}
\end{equation}
The classical momentum space distributions of the heavy quarks can be obtained by a Monte Carlo procedure as realized in PYTHIA ~\cite{Sjostrand:2006za}.  It can be verified that the momentum distribution of single heavy quarks agrees well with the QCD based FONLL calculations~\cite{Cacciari:2012ny,Song:2015sfa,Song:2015ykw}.

This procedure allows to calculate the distribution of the relative momentum $g^{\rm PYTHIA}_{c/b}(\mathbf{p})$ between the quarks and the antiquark of the pair. For the initial spatial separation of the heavy quark and heavy antiquark,  created at the same vertex, we assume a Gaussian distribution. This allows to write in a semi-classical picture the 2-body Wigner density for a 
$c\bar c$ or a $b\bar b$ pair produced in the same nucleon-nucleon collision as
\begin{equation}
W^{(2)}(\mathbf{r,p})
\sim r^2 \exp\bigg(-\frac{r^2}{2\delta^2}\bigg)g^{\rm PYTHIA}_{c/b}(\mathbf{p}),
\label{separation}
\end{equation}
neglecting $W^{(N-2)}$, which is mostly composed of light partons in pp collisions.
The spatial width is related to the inverse of heavy quark mass $m_Q$ by
$\delta^2=\langle r^2\rangle/3=4/(3m_Q^2)$. Consequently we find $\sqrt{\langle r^2\rangle}/2=1/m_Q$. To obtain a $W_{class}^{(2)}$
we randomly select the relative momentum and position from $W^{(2)}(\mathbf{r,p})$.
Now we have all elements to calculate the momentum distribution of a quarkonium of type $i$, that is,
Eq.~(\ref{eq:dnd3p}). The only quantities which enter the calculation are $\Phi_i$ and the mass of the heavy quark.

\section{$\Upsilon$ production in heavy-ion collisions}\label{AA}

As we saw already in section \ref{pp}, the probability to find a quarkonium is (neglecting all flavour color and spin factors) given by  
\begin{eqnarray}
P_\Phi = {\rm Tr}[\rho_\Phi \rho_N],
\label{density}
\end{eqnarray}
where $\rho_\Phi$ and $\rho_N$ are respectively the density operators of the quarkonium and of the $N$-body system, created in a heavy ion collisions. 
Because also here the quantal $\rho_N$ is not known we have to replace it as in the pp case by an average of classical phase space density distribution (Eq.~(\ref{eq:class})), given as
\begin{eqnarray}
W^{(N)}\approx \prod_{i=1}^N h^{3N}\delta(r_i-r_i^*(t))\delta(p_i-p_i^*(t)),
\label{deltas}
\end{eqnarray}
In our calculation $W^{(N)}$ is provided by  PHSD simulations that are consistent with experimental data on open heavy flavors as well as on light hadrons ~\cite{Song:2015sfa,Song:2015ykw,Song:2016rzw}.

 In a quantum system Eq.~(\ref {density})
gives also for large times the quarkonium probability but when we replace $\rho_N$ by a classical phase space density the quantum correlations are lost and asymptotically we have a system in which the particles
are far away from each other. Therefore, applied to large times, Eq.~(\ref{eq:ni}) gives a vanishing probability to form a quarkonium. To cope with this problem we do not calculate the projection at large times but the integral over the rate $\Gamma$:
\begin{equation}
    P(t\to\infty) = \int_0^\infty \frac{d P_\Phi}{dt}dt=\int_0^\infty \Gamma(t)dt.
\label{probability}    
\end{equation}

In the Remler formalism (which does not consider forces between the heavy quarks during the expansion) several rates contribute to the total rate $\Gamma$,
\begin{eqnarray}
\Gamma(t)=\Gamma_{\rm coll}(t)+\Gamma_{\rm local}(t)+\Gamma_{\rm diff}(t)\nonumber\\
= \sum_{i,j}\sum_{\nu_{i(j)}}  \int d^3r_1d^3p_1 ... d^3r_N d^3p_N(2\pi)^{3N}\nonumber\\
\times \Phi^W_{S/P}(r_i,p_i;r_j,p_j)
\bigg\{\delta\bigg(t-\nu_{i(j)}\bigg)W^{(N)}(t+\varepsilon)\nonumber\\
-\delta\bigg(t-(\nu-1)_{i(j)}\bigg)W^{(N)}(t+\varepsilon)\bigg\},~~~
\label{new}
\end{eqnarray}
where the first summation on the right hand side runs over all heavy (anti)quarks as in Eq.~(\ref{eq:ni}). $\nu_{i(j)}$ is the time of the $\nu$ th scattering of the heavy quark $i$ (or of the heavy antiquark $j$) with a QGP parton.
$\varepsilon$ in the $N$-body density matrix implies a slight time delay to take into account the momentum change of the heavy (anti)quark due to the scattering with a QGP parton at $t=\nu_{i(j)}$ or $t=(\nu-1)_{i(j)}$. 
The functions $\Phi^W_{S,P}$ will be defined in the next section. We see that, whenever a heavy (anti)quark scatters, the Wigner projection to the different quarkonium states, Eqs.~(\ref{wigner1}) and (\ref{wigner2}), are updated.  This projection is carried out for all possible combinations of heavy quark and heavy antiquark pairs, which involve the scattered heavy (anti)quark, but if they are far away  in coordinate or momentum space their contribution to the quarkonium yield is marginal.

Adding explanations on Eq.~(\ref{new}), 
the collisional rate, $\Gamma_{\rm coll}(t)$, describes the change of the probability to find a quarkonium due to collisions of one of its constituents with light partons of the QGP.  If the heavy quarks move on straight line trajectories between their collisions and not on curved ones due to their mutual interaction the original Remler formalism has to be supplemented by a diffusion rate, $\Gamma_{\rm diff}$ (t)~\cite{Song:2023ywt}.  The local rate, $\Gamma_{\rm local} (t)$, appears if the width of the Wigner density of the quarkonia depend on the temperature and hence on time~\cite{Villar:2021loy,Villar:2022sbv,Song:2023ywt}.  
The sum of the three rates is equivalent to the update of Wigner projection onto each quarkonium state, whenever a heavy (anti)quark scatters with a QGP parton~\cite{Song:2023ywt}, expressed by the right hand side of Eq.~(\ref{new}). One doesn't have to consider each rate separately.

Initially the temperature of the QGP is higher than the dissociation temperature, above
which the Schrödinger equation for the quarkonia does not give bound states.  The radii of the bound states as a function of the temperature are displayed in Fig.~\ref{radii}. Therefore the initially produced quarkonia do not contribute directly to the observed spectrum. Only when the QGP passes the dissociation temperature (which is different for the different quarkonia states) the production of quarkonia sets in. 
In our approach the probability to form a quarkonium does not change anymore when the expanding system has passed the hadronization temperature $T_c$. Then all light and heavy quarks form hadrons and are not available anymore
for the production of quarkonia.
For more details we refer to the Refs.~\cite{Villar:2021loy,Villar:2022sbv,Song:2023ywt}.
 We just want to point out that in this kinetic approach we follow all produced heavy quarks from their creation until the final appearance as part of a open or hidden heavy flavour hadron.  

For example, if there are only one heavy quark and one heavy antiquark and they experience two scatterings with a QGP parton, at $t=t_1$ and $t=t_2$, Eq.~(\ref{probability}) turns to
\begin{eqnarray}
P&=&\int_0^\infty \Gamma(t)dt=\widetilde{W}(t_0+\varepsilon)\nonumber\\
&&+ \widetilde{W}(t_1+\varepsilon)-\widetilde{W}(t_0+\varepsilon)\nonumber\\
&&+ \widetilde{W}(t_2+\varepsilon)-\widetilde{W}(t_1+\varepsilon)\nonumber\\
&=&\widetilde{W}(t_2+\varepsilon)
\label{example}
\end{eqnarray}
where $t_0$ 
is the time at the dissociation temperature and 
\begin{eqnarray}
\widetilde{W}(t_i)=(2\pi)^{6}\int d^3r_1d^3p_1 d^3r_2 d^3p_2\nonumber\\
\times \Phi^W_{S/P}(r_1,p_1;r_2,p_2)
W^{(2)}(t_i).
\end{eqnarray}
The first line in Eq.~(\ref{example}) indicates the initial formation at the dissociation temperature and the second and third lines show that the Wigner projection is updated at $t_1$ and at $t_2$. $\widetilde{W}(t_2+\varepsilon)$
signifies the last collision of the heavy (anti)quark before the QGP hadronizes.
As a result, only the Wigner projection at $t=t_2$ remains as shown in Eq.~(\ref{example}).


\section{Results}

To avoid the numerical integration of  Eq.~(\ref{eq:jpsiwign}) in the actual calculations one replaces
the Wigner density of the wave functions of the quarkonia by  simple harmonic oscillator (SHO) wave functions. Then the Wigner density of the $1 S-$ state and the $1 P-$ state are, respectively, given by~\cite{Baltz:1995tv,ExHIC:2011say},
\begin{eqnarray}
\Phi^W_{\rm S}({\bf r, p})&=&8\frac{D}{d_1 d_2}\exp\bigg[-\frac{r^2}{\sigma^2}-\sigma^2p^2\bigg],\label{wigner1}\\
\Phi^W_{\rm P}({\bf r, p})&=&\frac{16}{3}\frac{D}{d_1 d_2}\bigg(\frac{r^2}{\sigma^2}-\frac{3}{2}+\sigma^2p^2\bigg)\nonumber\\
&&\times\exp\bigg[-\frac{r^2}{\sigma^2}-\sigma^2p^2\bigg],
\label{wigner2}
\end{eqnarray}
where $\sigma^2=2/3\langle r^2\rangle$ for $S-$state and $\sigma^2=2/5\langle r^2\rangle$ for $P-$state with $\sqrt{\langle r^2\rangle}$ being the root-mean-square (rms) distance between the heavy quarks, and $D$, $d_1$, and $d_2$ are the color-spin degeneracy of quarkonium, quark and antiquark, respectively.
$\Phi^W_{\rm S}({\bf r, p})$ and $\Phi^W_{\rm P}({\bf r, p})$ are not a single eigen state of Eq.~(\ref{schrodinger1}). For example, $|\Phi_{\Upsilon(1S)}\rangle \langle \Phi_{\Upsilon(1S)}|$ in Eq.~(\ref{eq:jpsiwign}) is sum of three spin states,
$|S_z=1\rangle \langle S_z=1|$, $|S_z=0\rangle \langle S_z=0|$ and $|S_z=-1\rangle \langle S_z=-1|$. This gives rise to the factor
$D$ in Eqs.~(\ref{wigner1}) and (\ref{wigner2}).  In addition, in Eq.~(\ref{new}) one has to take into account that
there is only one possibility to combine the two color charges to a color singlet and the two spins to a desired $S_z$. This gives rise to the factor 1/($d_1$$d_2$). Associating the spin and color factor with $\Phi^W_{\rm S}({\bf r, p})$ we can drop them from Eqs.~(\ref{deltas}) and (\ref{new}).
The Wigner function for highly excited states such as $2 S-$, $2 P-$ and $3 S-$ states has a much more complicated form~\cite{Cho:2014xha,Kordell:2021prk}.
We assume that the simple forms of Eqs.~(\ref{wigner1}) and (\ref{wigner2}) is applicable to the excited states only by changing the parameter $\sigma$.

For the heavy quark production in pp collisions  we use the PHSD approach with the PYTHIA event generator ('in PHSD tune' \cite{Kireyeu:2020wou}) where the heavy quark production is additionally tuned in order to reproduce the FONLL transverse momentum distribution 
~\cite{Song:2015ykw}.

\subsection{proton-proton collisions}

\begin{figure}[h]
\centerline{
\includegraphics[width=9 cm]{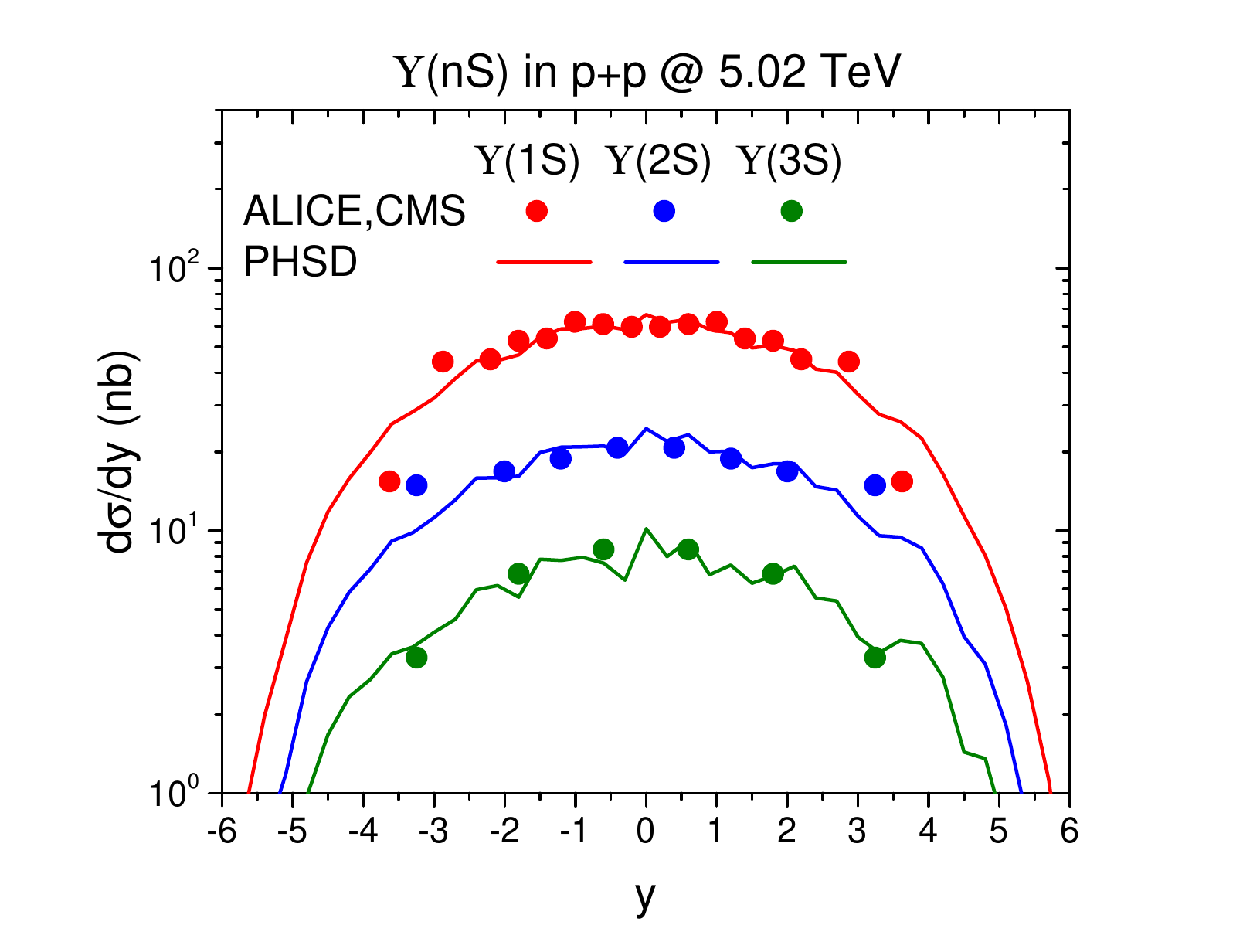}}
\centerline{
\includegraphics[width=9 cm]{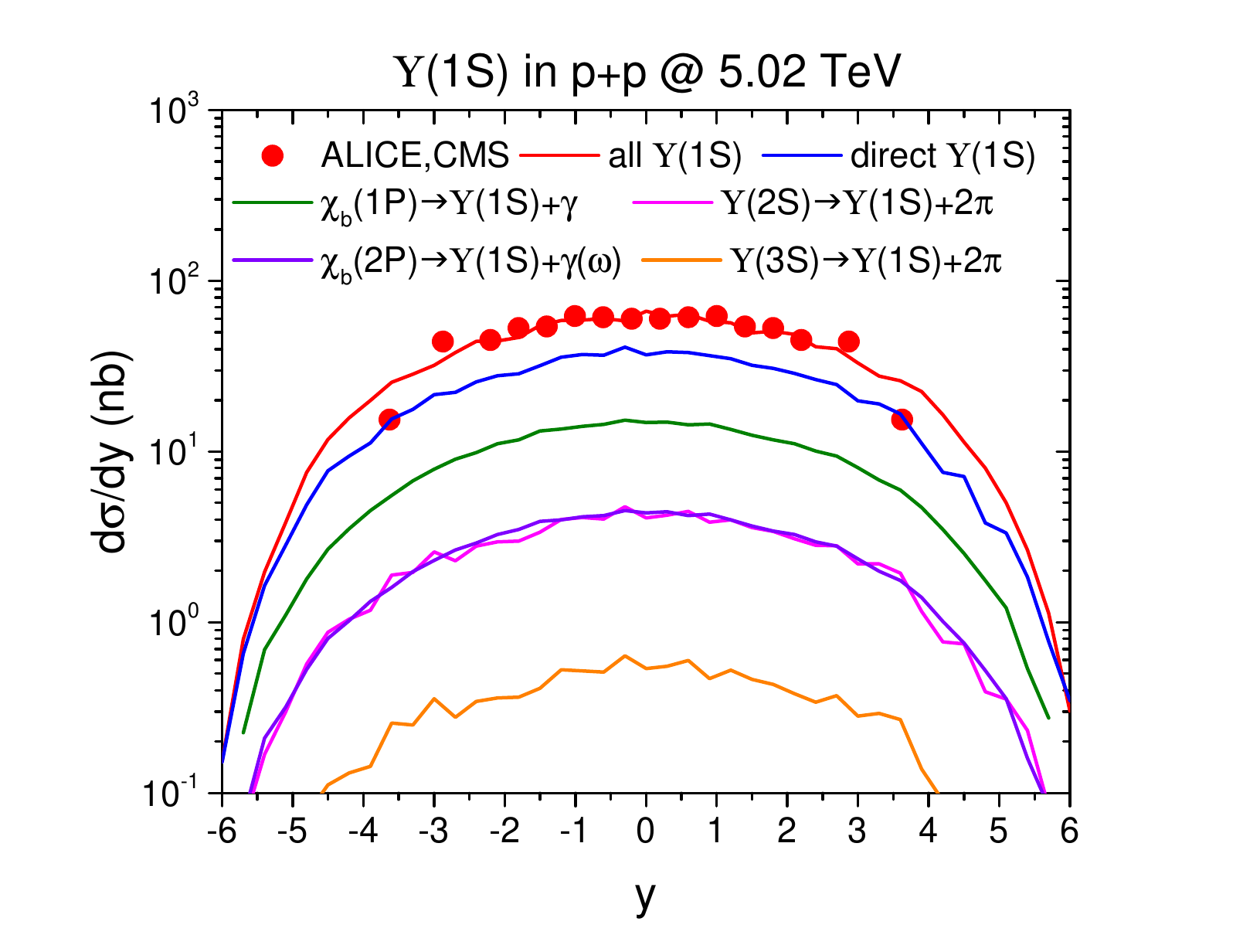}}
\caption{(Color online) (upper) The PHSD results for the rapidity distributions of inclusive $\Upsilon$ (1S), $\Upsilon$ (2S) and $\Upsilon$ (3S) and (lower) direct $\Upsilon$ (1S) and feed-down from the excited states to it in p+p collisions at $\sqrt{s_{\rm NN}}=$ 5.02 TeV for $r_{1S}=$ 0.22 fm, $r_{1P}=$ 0.22 fm, $r_{2S}=$ 0.3 fm, $r_{2P}=$ 0.24 fm and $r_{3S}=$ 0.38 fm, which are compared with the experimental data from the CMS and ALICE Collaboration~\cite{CMS:2018zza,ALICE:2021qlw}.}
\label{pp502y}
\end{figure}

Figure~\ref{pp502y} (top) shows the rapidity distribution of inclusive $\Upsilon$ (1S), $\Upsilon$ (2S) and $\Upsilon$ (3S) in p+p collisions at $\sqrt{s_{\rm NN}}=$ 5.02 TeV  and (bottom) the rapidity distribution of the direct $\Upsilon$ (1S)  as well as the feed-down contribution from the excited states. The radii, $\sqrt{\langle r^2\rangle}$/2, of $\Upsilon$ (1S), $\chi_b$ (1P), $\Upsilon$ (2S), $\chi_b$ (2P) and $\Upsilon$ (3S) are taken to be 0.22, 0.22, 0.3, 0.24 and 0.38 ${\rm fm}$, respectively~\cite{Song:2017phm}.
In the lower panel the contributions to the inclusive distribution from the direct $\Upsilon$ (1S) as well as  the feed-down from $\chi_b$ (1P), $\Upsilon$ (2S), $\chi_b$ (2P) and  $\Upsilon$ (3S) to the inclusive $\Upsilon$ (1S) are, respectively, 62 \%, 23 \%, 7 \%, 7 \% and 1 \%. This is consistent with the experimental data~\cite{CDF:1999fhr,LHCb:2012qym}. 

\begin{figure}[h]
\centerline{
\includegraphics[width=9 cm]{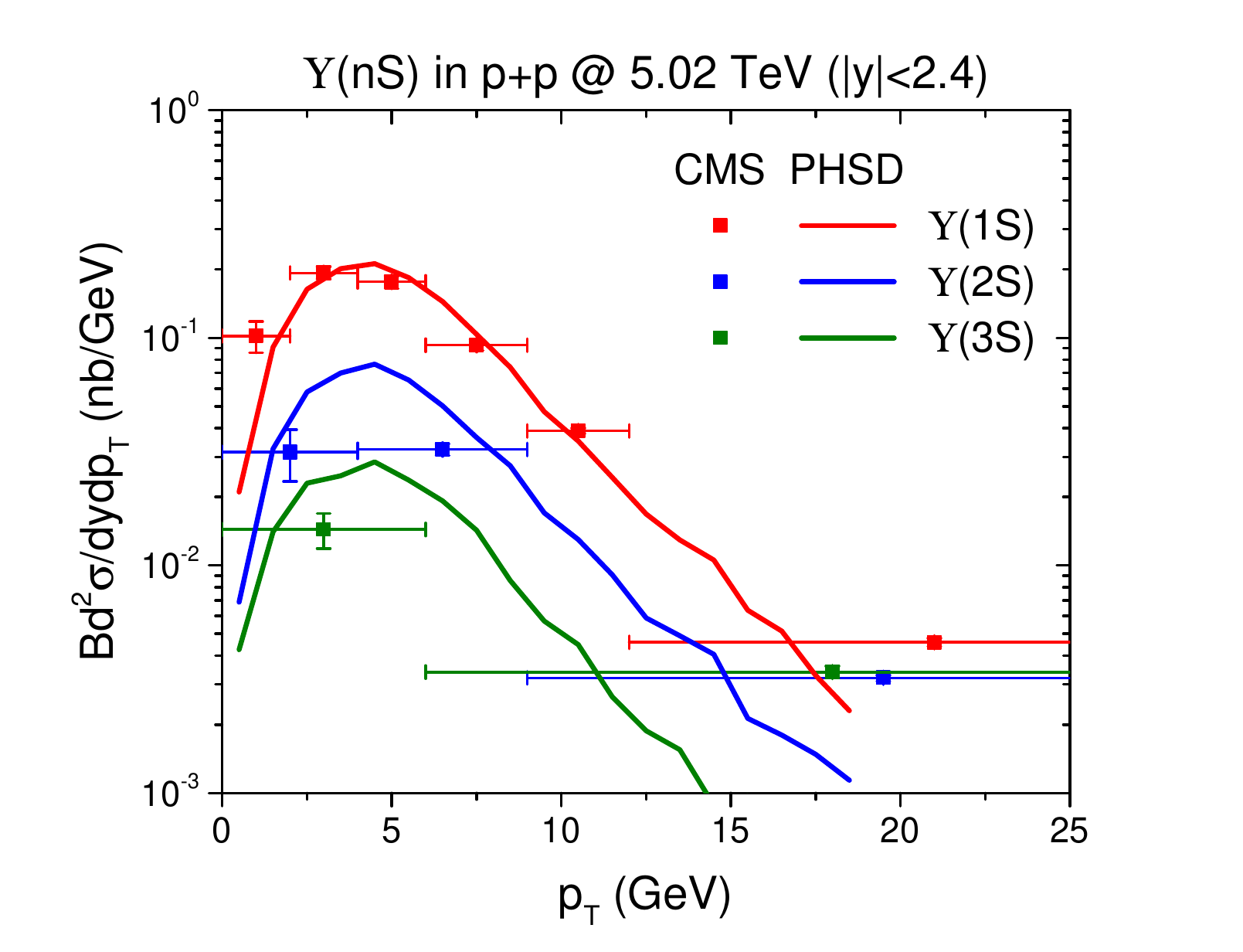}}
\centerline{
\includegraphics[width=9 cm]{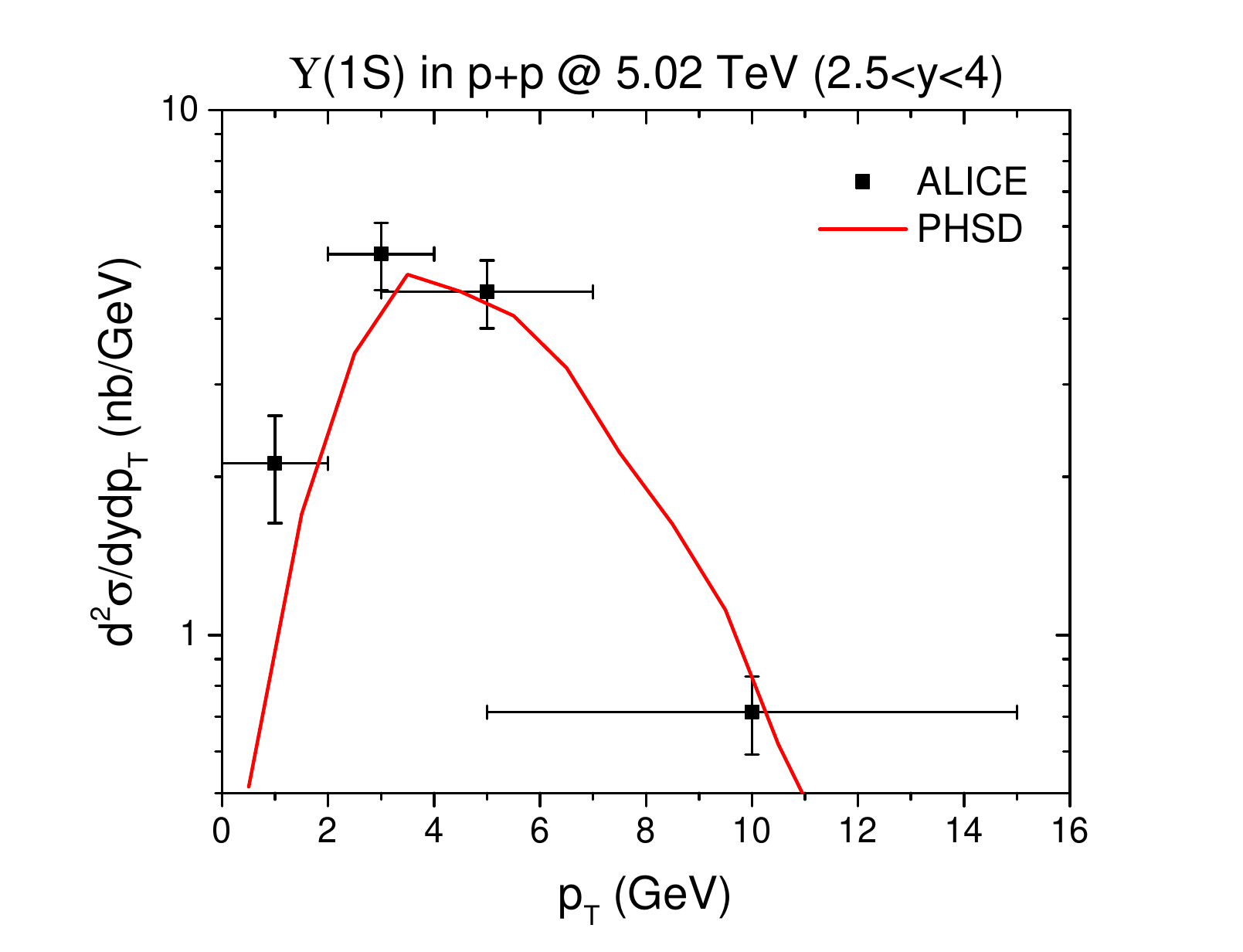}}
\caption{(Color online) (upper) The PHSD results for the transverse momentum spectra of $\Upsilon$ (1S), $\Upsilon$ (2S) and $\Upsilon$ (3S) in midrapidity and (lower) that of $\Upsilon$ (1S) in forward rapidity in p+p collisions at $\sqrt{s_{\rm NN}}=$ 5.02 TeV for $r_{1S}=$ 0.22 fm, $r_{1P}=$ 0.22 fm, $r_{2S}=$ 0.3 fm, $r_{2P}=$ 0.24 fm and $r_{3S}=$ 0.38 fm, which are compared with the experimental data from the CMS and ALICE Collaboration~\cite{CMS:2018zza,ALICE:2021qlw}.
}
\label{pp502pt}
\end{figure}

We display in Fig.~\ref{pp502pt} the $p_T$ spectra of inclusive $\Upsilon$ (1S), $\Upsilon$ (2S) and $\Upsilon$ (3S) in p+p collisions at the same energy. The upper panel displays the midrapidity results ($|y|<2.4)$, the lower panel those for  forward rapidity $2.5<y<4$. 

One can conclude that our approach reproduces the experimental data on bottomonium production in pp collisions not only in rapidity but also in transverse momentum.  Adding these results to those of Ref.~\cite{Song:2017phm} we can conclude that the density matrix approach reproduces well the measured rapidity and $p_T$ distribution of  all $c\bar c$ and $b \bar b$ quarkonium states.

\subsection{heavy-ion collisions}



\begin{figure}[h]
\centerline{
\includegraphics[width=9 cm]{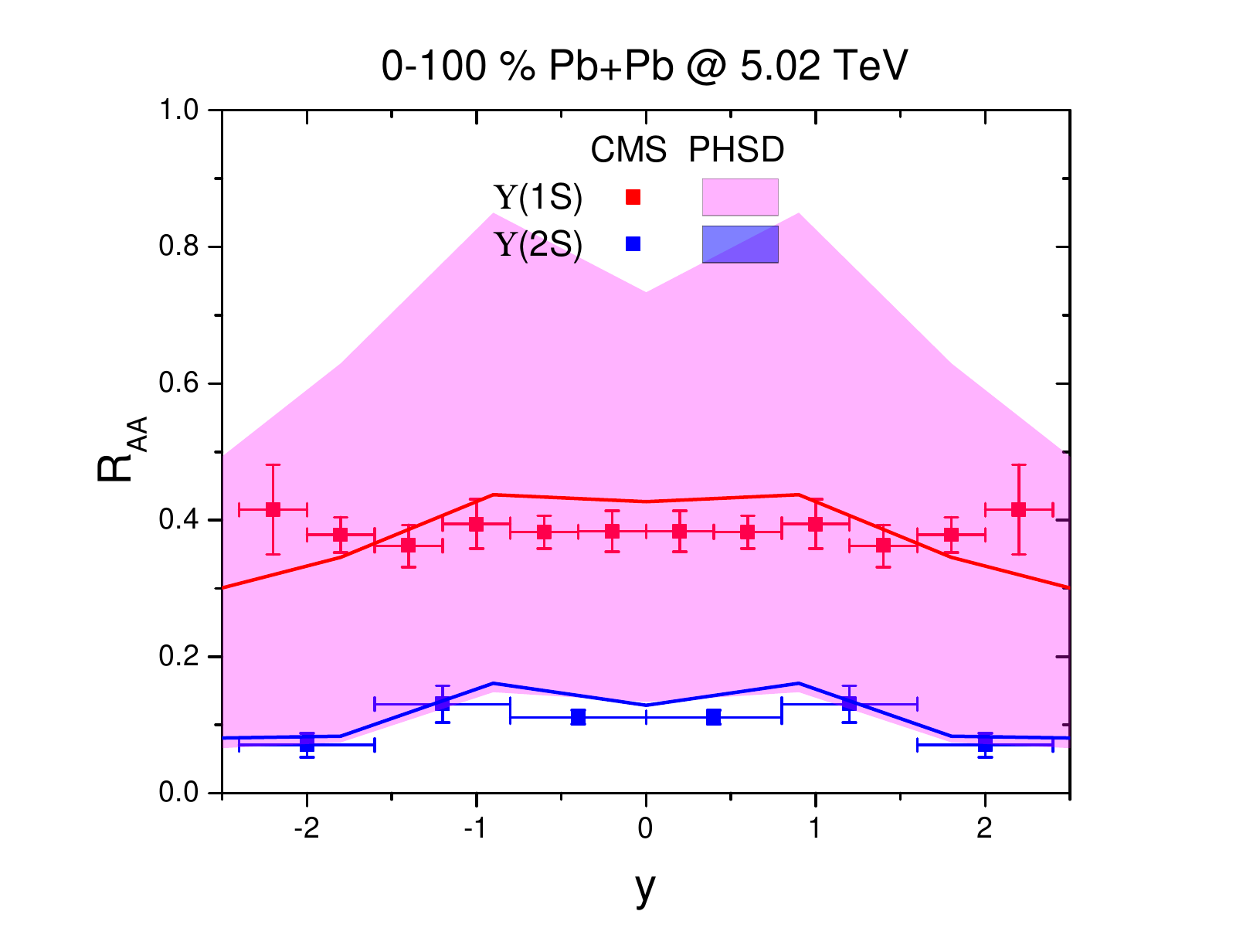}}
\centerline{
\includegraphics[width=9 cm]{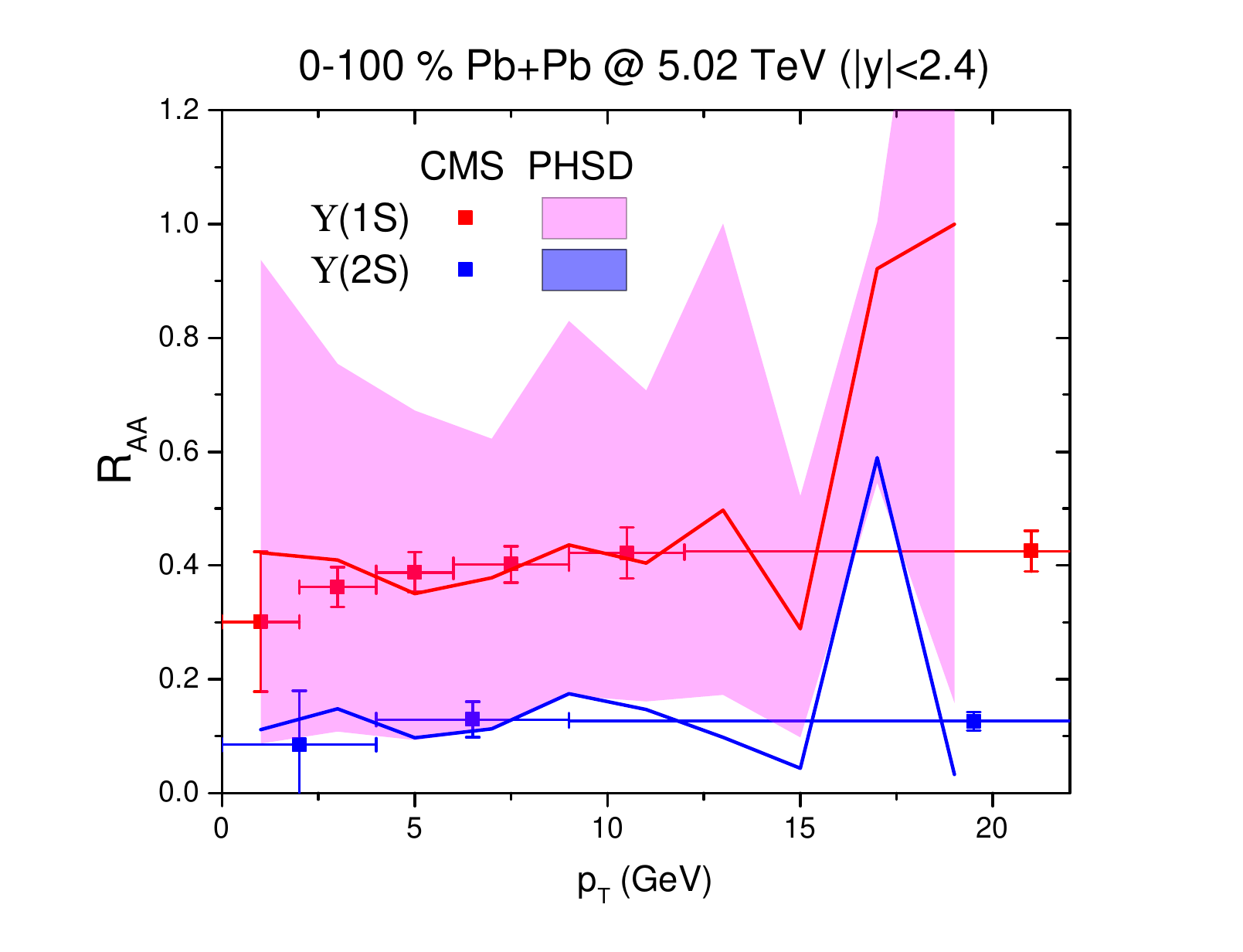}}
\caption{(Color online) The PHSD results for the $R_{\rm AA}$ of $\Upsilon$ (1S) and $\Upsilon$ (2S) for the free energy potential in 0-80 \% central Pb+Pb collisions at $\sqrt{s_{\rm NN}}=$ 5.02 TeV compared with experimental data in 0-100 \% central collisions from the CMS Collaboration~\cite{CMS:2018zza}.}
\label{raa}
\end{figure}

In heavy-ion collisions 
the production of heavy quarks is modified by the shadowing effects.
The parton distribution at small $x$ decreases and, as a result, the charm and bottom production is suppressed at small $p_T$, compared to pp collisions~\cite{Song:2015ykw}.
The shadowing effects are realized in PHSD by EPS09~\cite{Eskola:2009uj}.

Figure~\ref{raa} shows the nuclear modification factor $R_{AA}$ of $\Upsilon$ (1S) and $\Upsilon$ (2S) as a function of rapidity,  $R_{AA}= 
(dN^{PbPb}/dy)/(N_{coll}dN^{pp}/dy)$,
and of transverse momentum,  $R_{AA}= 
(dN^{PbPb}/dp_T)/(N_{coll}dN^{pp}/dp_T)$,
in 0-80 \% central Pb+Pb collisions at $\sqrt{s_{\rm NN}}=$ 5.02 TeV. $N_{coll}$ is the number of initial hard nucleon-nucleon collisions in a heavy-ion collision.

The upper limit of the band indicates the $R_{\rm AA}$ of $\Upsilon$ (1S) formed just below its dissociation temperature, corresponding to the first line in Eq.~(\ref{example}).  The yield includes the feed-down from excited states, which are also considered just below their dissociation temperatures. The lower limit of the band is $R_{AA}$ for the final $\Upsilon$ (1S), which have been formed when the expanding system passes the hadronization temperature $T_c$.  We see that the interaction with the QGP reduces $R_{AA}$ from roughly 0.7 to roughly 0.2.
For the $\Upsilon$ (2S) the band is much narrower due to its lower dissociation temperature, which is close to $T_c$.


In figure~\ref{raa} we display as well the experimental data from the CMS collaboration~\cite{CMS:2018zza}. 
First of all, one can see that the results of $\Upsilon(2S)$ are comparable to the experimental data. This agreement shows that if the bottomonium does not interact with the QGP,  the projection of the Wigner density of the $\Upsilon(2S)$ onto the Wigner density of the bottom (anti)quarks at $T_{diss}$ gives directly the experimental results, confirming our formalism.  For the $\Upsilon(1S)$ the experimental data are between the upper and lower limits of the band, both for the rapidity and the transverse momentum spectrum. This means that our approach shows too much suppression of the bottomonia during their passage through the QGP.

Before we discuss the possible origin of this observation we want to stress that, even without collisional interactions between heavy (anti)quarks and the QGP, we do not expect a $R_{AA}=1$.
Although we wait in AA collisions until the quarks are separated by the distance, given by the Monte Carlo method according to Eq.~(\ref{separation}), to apply formula (17) in order to be as close as possible to the pp collision, there are differences. 
Firstly if the quarkonium state is formed before the QGP is created, it has at the beginning of the expansion a temperature well above the dissociation temperature and so
the quarkonia will disintegrate. Secondly the quarkonium radius from heavy quark potential in a QGP is larger compared to the radius fitted to the experimental data in pp collisions, shown in Figs.~\ref{pp502y} and~\ref{pp502pt}.
Thirdly heavy quarks and antiquarks from different vertices can form a quarkonium.

One of the possible origins that our approach does not reproduce the heavy ion data may be the following:
If a bottom and an antibottom quark are close to each other in coordinate and momentum space, where the probability that they form a color singlet state is large,  the scattering cross section of its constituents with a thermal parton will be suppressed.  Soft gluons do not see then anymore the individual color charges
but only a color singlet. Therefore the scattering rate of such a narrow state is lower and so is the probability that it falls apart into a $b$ and a $\bar b$ quark. We therefore made calculations in which the scattering rate of $b$ and $\bar b$ with QGP partons is reduced arbitrarily by 90 \%, motivated to get a reasonable agreement with data.
In Eq.~(\ref{example}) this reduction is realized by executing the scatterings at $t_1$ and $t_2$ with 10 \% probability by Monte Carlo for the update of Wigner projection. 

The result is presented as red and blue solid line for $\Upsilon$ (1S) and $\Upsilon$ (2S) in Fig.~\ref{raa}.
One can see that with such a reduction our results agree with the experimental data from the CMS Collaboration~\cite{CMS:2018zza}.

Such a simplistic reduction is only a very crude approximation. In reality the reduction should depend
on the quarkonia states, due to their different radii, as well as on the momentum transfer, in other words, the wave length of the exchanged gluon.  This will be explored in an upcoming publication. 
In the present study we assume the same reduction factor for the ground state as well for excited states.
Since the dissociation temperature of the excited state is close to $T_c$, the reduction of heavy quark scattering does not influence $R_{AA}$ of $\Upsilon(2S)$ (thin blue lines in Fig.~\ref{raa}).

If the heavy quark potential is stronger than the free energy, the upper limit of the band will be higher, because bottomonium is formed earlier, while the lower limit of the band will stay similar. Then, considering the upper limit corresponds to the complete suppression of scattering and the lower limit to no suppression, the interaction rate for the stronger potential should be less suppressed 
to reproduce the experimental data in Fig.~\ref{raa}.

\begin{figure}[h]
\centerline{
\includegraphics[width=9 cm]{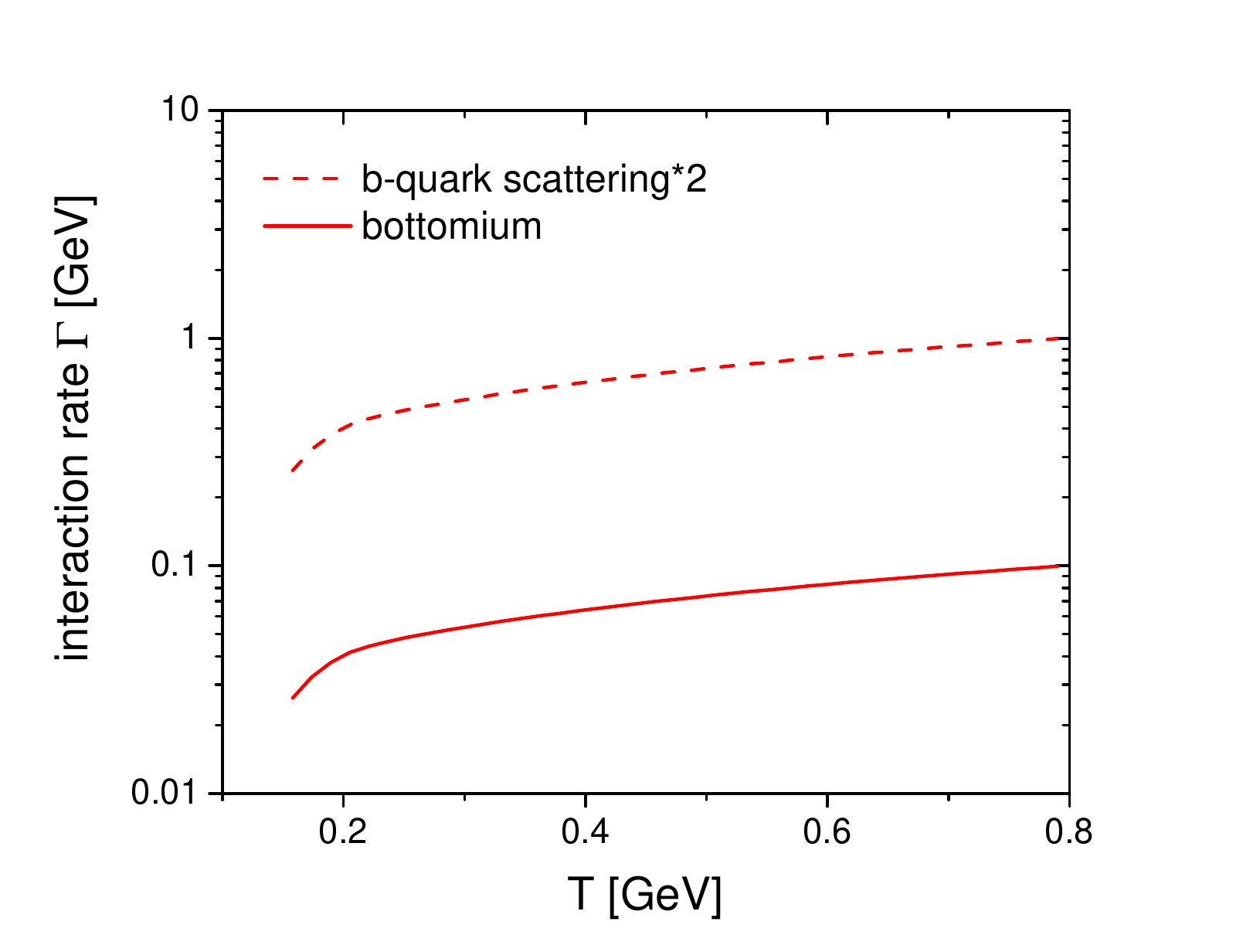}}
\centerline{
\includegraphics[width=9 cm]{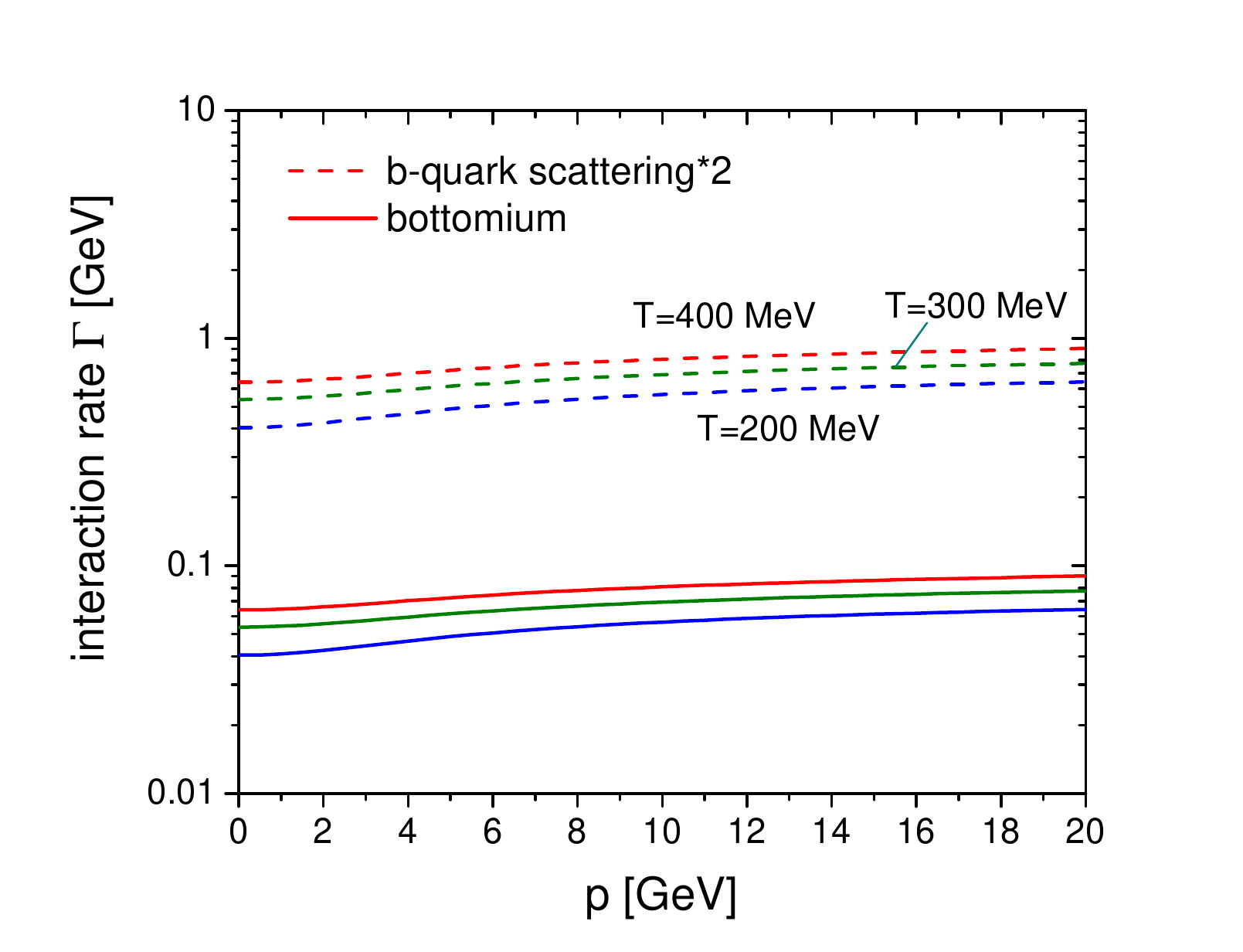}}
\caption{(Color online) The PHSD results for the  reaction rate of $\Upsilon$ as a function of temperature and momentum.}
\label{rate}
\end{figure}
Figure~\ref{rate} shows the interaction rate of $\Upsilon$, which is 10 \% of the interaction rate of open bottom and antibottom quarks, as a function of temperature (top) and momentum (bottom) in a QGP within the dynamical quasi-particle model~\cite{Berrehrah:2013mua,Moreau:2019vhw}. 
The dashed lines show the results if one doubles the reaction rate of bottom quarks while the solid lines are the result if one reduces the rates by a factor of 10. 
One can see that the reduced rates are less than 100 MeV. The interaction rate of bottomonium 
does not necessarily mean the dissociation rate of bottomonium which is widely used in many phenomenological models for bottomonium dissociation in heavy-ion collisions, because the rate in Fig.~\ref{rate} (see Eq. (\ref{new})) includes both dissociation and regeneration and  depends on how the momentum change of bottom (anti)quark, induced by the interaction, affects the projection of the Wigner density 
on  the bottomonium  Eq.~(\ref{example}). If the interaction of a bottom quark or a bottom antiquark does not change the expectation value, it can be interpreted as a elastic scattering of bottomonium, though it will rarely happen. In the case of an expanding QGP the interactions 
have the tendency to lower the multiplicity of bottomonium states. So the interaction rate is closely related to the dissociation rate in other models. We note in passing that the  absorption of time-like gluons, i.e. $\Upsilon+g \to b+\bar b$ \cite{Peskin:1979va}, does not exist in this approach.


\begin{figure}[h]
\centerline{
\includegraphics[width=9 cm]{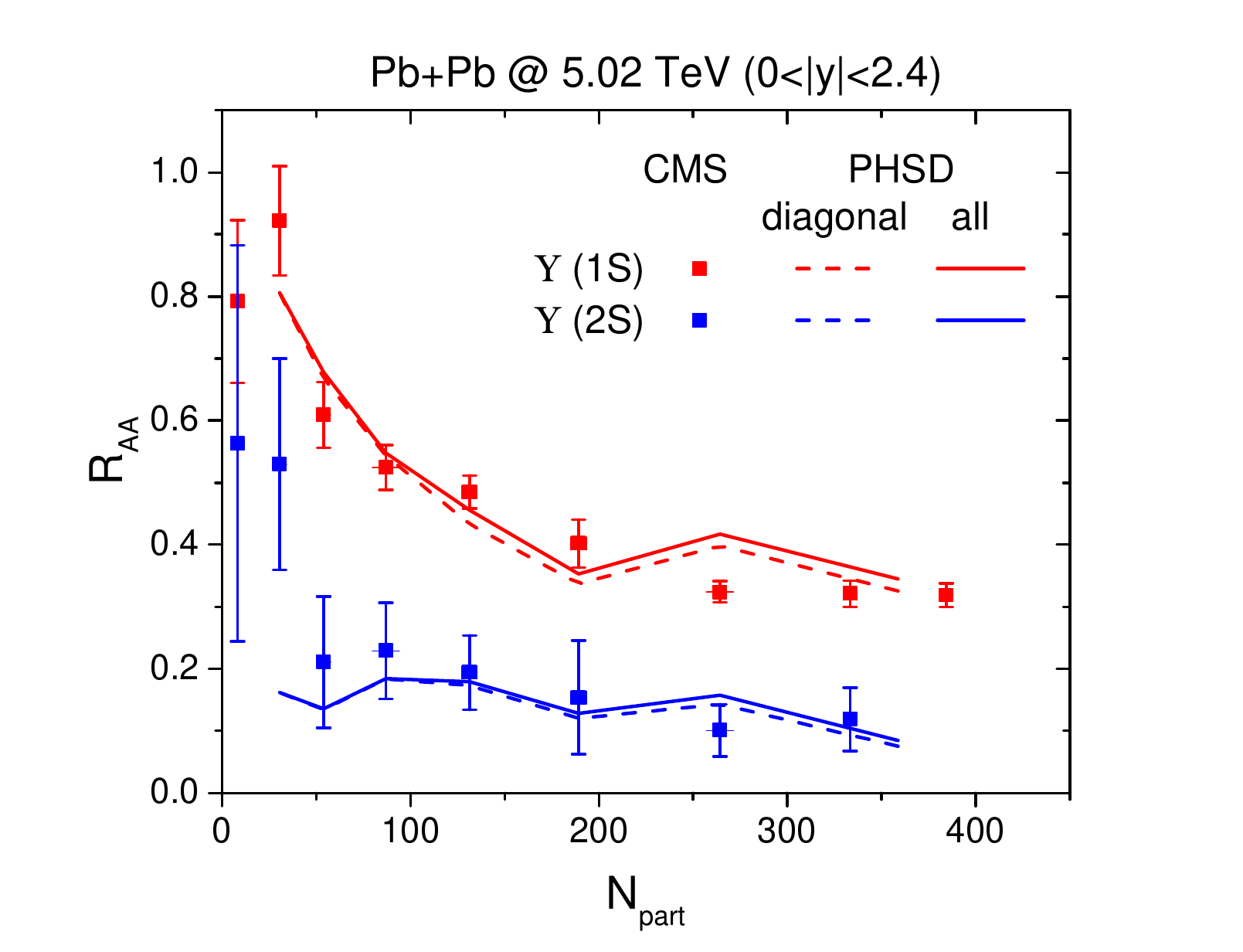}}
\caption{(Color online) The PHSD results for the  $R_{\rm AA}$ of $\Upsilon$ (1S) and $\Upsilon$ (2S) as a function of the number of participants in Pb+Pb collisions at $\sqrt{s_{\rm NN}}=$ 5.02 TeV, compared with experimental data from the CMS Collaboration~\cite{CMS:2018zza}. The solid lines are full results while the dashed lines only from the coalescence of the original bottom pairs. }

\label{npart}
\end{figure}
Last we display in Fig.~\ref{npart} the $R_{\rm AA}$ of $\Upsilon$ (1S) and $\Upsilon$ (2S) as a function of the number of participants in Pb+Pb collisions at $\sqrt{s_{\rm NN}}=$ 5.02 TeV.
The solid lines are the results from all bottomonia and the dashed lines are those  from the diagonal bottomonia only where bottomonia with a $b$ and a $\bar b$ from two different initial vertices are excluded.
One can see that the off-diagonal contribution (in contradistinction to the charmonium production) is not important because even at LHC energies only a small number of bottom quarks is produced.


\section{summary}\label{summary}

In this study we have extended our previous study on charmonium production in proton-proton collisions to bottomonia. 
As for the charmonia the bottomonia production is factorized into the production of the $b\bar b$ pair, described perturbatively, 
and a non-perturbative formation of $\Upsilon$ (1S) and $\Upsilon$ (2S). The first step is realized by the PYTHIA event generator.
For the latter we solved the Schrödinger equation for the quarkonia employing a potential calculated by lattice QCD. From
the eigenstates we can obtain the Wigner density. The projection of the $N$-body Wigner density of the system on the Wigner densities
of the different quarkonium eigenstates gives the probability that such a quarkonium is formed. 
The only inputs in this calculation are the eigen functions of the quarkonium wave functions and the mass of the heavy quark.
In this approach the experimental rapidity distribution as well as the experimental transverse momentum distribution of all quarkonia states
can be well reproduced and establishes this method as a general approach for quarkonia production at RHIC and LHC energies.

In a heavy ion collisions in which a QGP is formed, the bottomonia which are formed in the initial hard collisions do not survive,
because beyond a temperature of around 3.3 (1.2) $T_c$ the $\Upsilon$ (1S)($\Upsilon$ (2S)) is not bound anymore, unless $b\bar{b}$ pair is produced in a relatively cold region of QGP.
During the expansion the heavy quarks interact with the partons of the QGP, as described by PHSD, and can form bound states, when the temperature falls below $T_{diss}$ until the QGP hadronizes. 
We note that the dissociation temperature of quarkonium can be slightly  lower for a large imaginary potential and/or a large thermal width~\cite{Song:2020kka}.

The final yield of the $\Upsilon$ (1S) state is too small, but agrees with experimental data when the heavy quark-parton scattering cross section is reduced to 10 \%.
This is a hint that indeed this cross section is reduced if the heavy quark is part of a bottomonium, as expected because soft gluons  see only a color neutral object. To investigate this in detail is the subject of an upcoming publication. We also expect that the inclusion of a heavy quark potential for the dynamics of heavy (anti)quark will improve our results~\cite{Villar:2022sbv}.


\section*{Acknowledgements}
We acknowledge support by the Deutsche Forschungsgemeinschaft (DFG, German Research Foundation) through the grant CRC-TR 211 'Strong-interaction matter under extreme conditions' - Project number 315477589 - TRR 211. 
This work is supported by the European Union’s Horizon 2020 research and innovation program under grant agreement No 824093 (STRONG-2020).
The computational resources have been provided by the LOEWE-Center for Scientific Computing and the "Green Cube" at GSI, Darmstadt and by the Center for Scientific Computing (CSC) of the Goethe University.

\bibliography{upsilon}

\end{document}